\documentclass[aps,prb,twocolumn,showpacs,preprintnumbers,amsmath,amssymb,superscriptaddress]{revtex4}
\usepackage{mathptm}  
\usepackage{dcolumn}                    
\usepackage{bm}                        
\usepackage{graphicx}
\usepackage{times}
\usepackage{epstopdf}
\usepackage{color}
\begin{document}

\definecolor{Red}{rgb}{1,0,0}

\definecolor{Blu}{rgb}{0,0,01}

\definecolor{Green}{rgb}{0,1,0}

\newcommand{\red}{\color{Red}}
\newcommand{\blu}{\color{Blu}}
\newcommand{\green}{\color{Green}}

\newcommand{\Imag}{{\Im\mathrm{m}}}   
\newcommand{\Real}{{\mathrm{Re}}}   
\newcommand{\im}{\mathrm{i}}        
\newcommand{\talpha}{\tilde{\alpha}}
\newcommand{\ve}[1]{{\mathbf{#1}}}

\newcommand{\x}{\lambda}  
\newcommand{\y}{\rho}     
\newcommand{\T}{\mathrm{T}}   
\newcommand{\Pv}{\mathcal{P}} 
\newcommand{\vk}{\ve{k}} 
\newcommand{\vp}{\ve{p}} 

\newcommand{\vm}{\boldsymbol{m}} 
\newcommand{\vM}{\boldsymbol{M}}
\newcommand{\X}{\mathcal{X}}
\newcommand{\Hp}{\mathcal{H}_\perp}

\newcommand{\N}{\underline{\mathcal{N}}} 
\newcommand{\Nt}{\underline{\tilde{\mathcal{N}}}} 
\newcommand{\g}{\underline{\gamma}} 
\newcommand{\gt}{\underline{\tilde{\gamma}}} 

\newcommand{\jacob}[1]{\textcolor{red}{#1}}

\newcommand{\vecr}{\ve{r}} 
\newcommand{\vq}{\ve{q}} 
\newcommand{\ca}[2][]{c_{#2}^{\vphantom{\dagger}#1}} 
\newcommand{\cc}[2][]{c_{#2}^{{\dagger}#1}}          
\newcommand{\da}[2][]{d_{#2}^{\vphantom{\dagger}#1}} 
\newcommand{\dc}[2][]{d_{#2}^{{\dagger}#1}}          
\newcommand{\ga}[2][]{\gamma_{#2}^{\vphantom{\dagger}#1}} 
\newcommand{\gc}[2][]{\gamma_{#2}^{{\dagger}#1}}          
\newcommand{\ea}[2][]{\eta_{#2}^{\vphantom{\dagger}#1}} 
\newcommand{\ec}[2][]{\eta_{#2}^{{\dagger}#1}}          
\newcommand{\su}{\uparrow}    
\newcommand{\sd}{\downarrow}  
\newcommand{\Tkp}[1]{T_{\vk\vp#1}}  
\newcommand{\muone}{\mu^{(1)}}      
\newcommand{\mutwo}{\mu^{(2)}}      
\newcommand{\epsk}{\varepsilon_\vk}
\newcommand{\epsp}{\varepsilon_\vp}
\newcommand{\e}[1]{\mathrm{e}^{#1}}
\newcommand{\dif}{\mathrm{d}} 
\newcommand{\diff}[2]{\frac{\dif #1}{\dif #2}}
\newcommand{\pdiff}[2]{\frac{\partial #1}{\partial #2}}
\newcommand{\mean}[1]{\langle#1\rangle}
\newcommand{\abs}[1]{|#1|}
\newcommand{\abss}[1]{|#1|^2}
\newcommand{\Sk}[1][\vk]{\ve{S}_{#1}}
\newcommand{\pauli}[1][\alpha\beta]{\boldsymbol{\sigma}_{#1}^{\vphantom{\dagger}}}

\newcommand{\eq}{Eq.}
\newcommand{\eqs}{Eqs.}
\newcommand{\cf}{\textit{cf. }}
\newcommand{\ie}{\textit{i.e. }}
\newcommand{\eg}{\textit{e.g. }}
\newcommand{\etal}{\emph{et al.}}
\def\i{\mathrm{i}}

\title{Pressure-Induced $0-\pi$ Transitions and Supercurrent Crossover in Antiferromagnetic Weak Links}

\author{Henrik Enoksen}
\affiliation{Department of Physics, Norwegian University of
Science and Technology, N-7491 Trondheim, Norway}

\author{Jacob Linder}
\affiliation{Department of Physics, Norwegian University of
Science and Technology, N-7491 Trondheim, Norway}

\author{Asle Sudb{\o}}
\affiliation{Department of Physics, Norwegian University of
Science and Technology, N-7491 Trondheim, Norway}

\date{\today}

\begin{abstract}
We  compute self-consistently the Josephson current in a superconductor-antiferromagnet-superconductor junction using a lattice model, focusing on $0-\pi$ transitions occurring 
when the width of the antiferromagnetic region changes from an even to an odd number of lattice sites. {Previous studies predicted $0-\pi$ transitions when alternating between an 
even and an odd number of sites for sufficiently strong antiferromagnetic order. We study numerically the magnitude of the threshold value for this to occur, and also explain the 
physics behind its existence in terms of the phase-shifts picked up by the quasiparticles constituting the supercurrent in the antiferromagnet.} Moreover, we show that this threshold 
value allows for \textit{pressure-induced $0-\pi$ transitions} by destroying the antiferromagnetic nesting properties of the Fermi surface, a phenomenon which has no counterpart in
ferromagnetic Josephson junctions, {offering a new way to tune the quantum ground state of a Josephson junction without the need of multiple samples.}
\end{abstract}

\maketitle

\textit{Introduction}. The study of the interplay between superconductivity and magnetism has been of considerable interest in condensed matter physics over the last decades. 
Phenomena such as the $0-\pi$-transition~\cite{ryazanov_prl_01} in ferromagnetic Josephson junctions has received much attention both from a fundamental quantum physics point of 
view in addition to being suggested as a potential basis for qubits.~\cite{terzioglu_apls_98}. While most of the focus in the above context has been on ferromagnetic (F) order, antiferromagnetic (AF) Josephson junctions are also of fundamental interest, due to the close relationship between the superconducting (S) phase and the antiferromagnetic phase 
in for instance high-temperature cuprate and iron-pnictide superconductors. Superconductivity and antiferromagnetism spin-density wave states may even coexist in the superconducting 
pnictides.~\cite{mazin_nature_10} Similar to SFS junctions, antiferromagnetic Josephson junctions (SAFS) have been predicted to display $0-\pi$-transitions \cite{andersen_prl_06}. 
However, for SAFS these transitions display a high sensitivity to the exact number of atomic layers (even vs. odd number) in the antiferromagnet. Ref.~\onlinecite{andersen_prb_08} reported that an antiferromagnetic Josephson junction is in a $\pi$-state for an odd number of layers, while it is in the $0$-state for an even number of layers provided that the 
antiferromagnetic order is much stronger than the superconducting order. An even-odd effect has also been observed in Josephson junctions with magnetic impurities in the 
middle layer.~\cite{wang_physica_C_12} 

In this Rapid Communication, we report on a novel aspect of antiferromagnetic Josephson junctions which allows for control over $0-\pi$ transitions within \textit{a single sample} 
in a way which has no counterpart in SFS structures. {We first compute numerically the threshold value for the antiferromagnetic order parameter at which the even-odd effect occurs.} 
Below this threshold, even and odd junctions behave qualitatively similar, both displaying a monotonic decay of the supercurrent with  superimposed small-scale oscillations, but 
without any sign-change of the critical current. As a result of this, we show that it is possible to obtain \textit{pressure-induced} $0-\pi$ transitions in antiferromagnetic 
Josephson junctions. Namely, applying pressure alters the Fermi level by moving it away from the van Hove singularity of the system, which simultaneously destroys the Fermi surface 
nesting properties giving antiferromagnetism in the first place. In this way, the pressure controls the magnitude of the staggered order parameter which triggers a $0-\pi$ transition 
once it drops below the aforementioned threshold value. This effect has no counterpart in conventional SFS junctions, since ferromagnetism does not rely on Fermi-surface nesting. 
The antiferromagnetic order thus offers a novel mechanism for controlling the quantum mechanical ground state of a Josephson junction.

\begin{figure}
  \centering
  \includegraphics[width=\columnwidth]{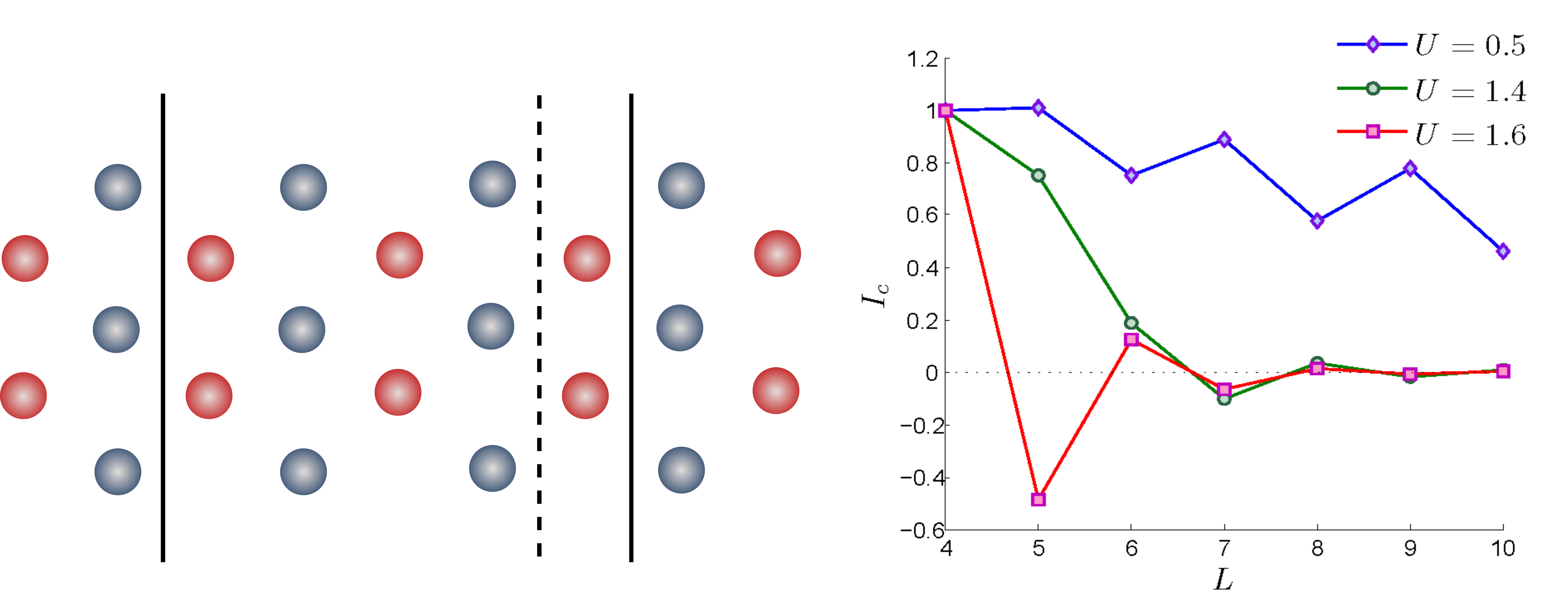}
  \caption{\textit{Left panel}: A model of the Josephson junction in (110)-direction. Red (blue) dots denote the A (B)-lattice. The dashed vertical line is the interface for an even junction, while the right solid line is the interface for an odd junction. The left solid line is the same for both even and odd junctions. \textit{Right panel}: Critical current as a function of length $L$ for different values of the magnetic coupling constant $U$. The data are normalized to the $L = 4$ value to simplify comparison. We have used $V = 1.9$.}
  \label{fig:lattice}
\end{figure}

\textit{Theory}.
Our approach closely follows that of Ref. \onlinecite{andersen_prb_05}. The system in question consists of an itinerant antiferromagnet sandwiched between two conventional 
$s$-wave superconductors. The interfaces are cut in the (110) direction and are considered transparent. We model the antiferromagnetic region as consisting of two square 
sublattices shown in Fig.~\ref{fig:lattice}, where the A- and B-lattice have oppositely preferred spin directions. The mean-field Hamiltonian then reads (see for instance 
Ref. \onlinecite{andersen_prl_06})
\begin{align}
  \hat{H} &= -t\sum_{\langle ij \rangle\sigma}{\hat{c}_{i\sigma}^{\dagger}\hat{c}_{j\sigma}} + \sum_{i}{\left(\Delta_i \hat{c}_{i\uparrow}^{\dagger}\hat{c}_{i\downarrow}^{\dagger} + \mathrm{H.c.}\right)} - \mu\sum_{i\sigma}{{\hat{c}_{i\sigma}^{\dagger}\hat{c}_{i\sigma}}} \nonumber \\
  &+ \sum_{i}{m_i\left(\hat{c}_{i\uparrow}^{\dagger}\hat{c}_{i\uparrow} - \hat{c}_{i\downarrow}^{\dagger}\hat{c}_{i\downarrow} \right)},
  \label{eq:hamiltonian}
\end{align}
where $\hat{c}_{i\sigma}^{\dagger}$ creates an electron with spin $\sigma$ on lattice site $i$, $\mu$ is the chemical potential, $t$ denotes the nearest-neighbor hopping integral, 
and $m_i$ and $\Delta_i$ are the magnetic and superconducting order parameters, respectively.

We use the Bogoliubov-Valatin transformations $c_{i\sigma} = \sum_n {\left(u_{n\sigma}(i) \gamma_{n\sigma} - \sigma v^*_{n\sigma}(i)\gamma^{\dagger}_{n\bar{\sigma}} \right)}$, where $\bar{\sigma} = -\sigma$, to obtain the standard Bogoliubov-de Gennes equations (BdG) \cite{andersen_prb_05}
\begin{equation}
    \sum_j \begin{pmatrix}
    H_{ij\sigma} & \Delta_{ij\sigma} \\
    \Delta_{ji\sigma}^* & -H_{ij\sigma}^*  
  \end{pmatrix} \begin{pmatrix}
    u_{n\sigma}(j) \\
    v_{n\bar{\sigma}}(j)
  \end{pmatrix} = E_n \begin{pmatrix}
    u_{n\sigma}(i) \\
    v_{n\bar{\sigma}}(i)
  \end{pmatrix}.
  \label{eq:bdg}
\end{equation}
Here, $H_{ij}\sigma = -t\delta_{\langle i,j \rangle} - \mu\delta_{ij} + \sigma m_i \delta_{ij}$ where $\sigma = \pm 1$ for spin up and down, and $\delta_{ij}$ and $\delta_{\langle i,j \rangle}$ are the Kroenecker deltas for onsite and nearest-neighbor sites, respectively. The off-diagonal block $\Delta_{ij\sigma} = -\Delta_i \delta_{ij}$ for $s$-wave symmetry.

Due to the crystal periodicity along the interface, we can Fourier transform the BdG-equations to make the problem effectively one-dimensional. Using the method described 
in Ref.~\onlinecite{andersen_prb_05}, we get the one-dimensional BdG-equations for the (110)-direction \cite{andersen_prb_05}
  \begin{align}
    &-\mu u_{i,\sigma}^{A(B)}(k_y) - 2t\cos{\frac{k_y}{\sqrt{2}}}e^{\pm i k_y/\sqrt{2}}\left[u_{i,\sigma}^{B(A)}(k_y) + u_{i \mp 1,\sigma}^{B(A)}(k_y) \right] \notag\\
    &+ \sigma m_i^{A(B)}u_{i,\sigma}^{A(B)}(k_y) - \Delta_{j}^{A(B)}v_{i,\bar{\sigma}}^{A(B)}(k_y) = E u_{i,\sigma}^{A(B)}(k_y), \label{eq:latticebdgu} \\
    &\mu v_{i,\bar{\sigma}}^{A(B)}(k_y) + 2t\cos{\frac{k_y}{\sqrt{2}}}e^{\pm i k_y/\sqrt{2}}\left[v_{i,\bar{\sigma}}^{B(A)}(k_y) + v_{i \mp 1,\bar{\sigma}}^{B(A)}(k_y) \right] \notag\\
    &+ \sigma m_i^{A(B)}v_{i,\bar{\sigma}}^{A(B)}(k_y) - {\Delta_{j}^{A(B)}}^*u_{i,\sigma}^{A(B)}(k_y) = E v_{i,\bar{\sigma}}^{A(B)}(k_y). \label{eq:latticebdgv}
  \end{align}
The $A(B)$ denotes the A(B) sublattice while the order parameters are defined as $\Delta_i = -V_i \langle \hat{c}_{i\downarrow}\hat{c}_{i\uparrow} \rangle$ and $m_i^{A(B)} = (U_i /2)(\langle \hat{n}_{i\uparrow}^{A(B)}\rangle - \langle \hat{n}_{i\downarrow}^{A(B)}\rangle)$. These are determined self-consistently through \cite{andersen_prb_05}
\begin{align}
  n_{i\sigma}^{A(B)} &= \sum_{n,k_y}\left[|u_{n,i,\sigma}^{A(B)}(k_y)|^2 f(E_{n,k_y,\sigma})\right. \nonumber \\ 
    &+ \left.|v_{n,i,\sigma}^{A(B)}(k_y)|^2 f(-E_{n,k_y,\sigma})\right],    \label{eq:SCmagn} \\
  \Delta_i^{A(B)} &= -V_i \sum_{n,k_y,\sigma}{u_{n,i,\sigma}(k_y)v_{n,i,\bar{\sigma}}^*(k_y) \tanh\left(\frac{\beta E_{n,k_y,\sigma}}{2}\right) }, \label{eq:SCdelta}
\end{align}
where $f(E) = (1 + \exp{(\beta E)})^{-1}$ is the Fermi-Dirac distribution. Here, $U_i$ and $V_i$ denote the spatially dependent coupling constants for the magnetization and 
superconducting pairing, respectively. We model the coupling contants as $U_i  = U\Theta(i-i_L)\Theta(i_R-i)$, $V_i  = V\Theta(i_L-i) + V\Theta(i-i_R)$, where $\Theta(x)$ is 
the Heaviside step function, and $i_{L,R}$ denote the left and right interface coordinates, respectively.

The dc Josephson current is obtained from \cite{andersen_prl_06,andersen_prb_08}
\begin{align}
 J &= -\frac{8te}{\hbar N_k}\sum_{n,k_y,\sigma} \cos\left(\frac{k_y}{\sqrt{2}}\right)\left[\mathrm{Im}( u_{n\sigma}(x)u_{n,\sigma}^*(x'))f(E_{n\sigma})\right. \nonumber \\
   &+ \left.\mathrm{Im}( v_{n\sigma}(x)v_{n,\sigma}^*(x'))f(-E_{n\sigma}) \right], \label{eq:Jcurrent}
\end{align}
where $e$ is the electron charge, and $x$ and $x'$ denote neighboring points in the antiferromagnet.

We fix the phase at the end of each superconductor, obtain self-consistent solutions from the above equations, and use Eq.~\eqref{eq:Jcurrent} to calculate the Josephson current 
for that particular phase difference. We set $\mu = 0$ throughout, and $t$ is used as the unit of energy in all calculations.

\textit{Results and Discussion}. In the right panel of Fig. \ref{fig:lattice} we show results for the critical current as a function of the number of AF-layers in the junction, 
for various values of the magnetization coupling constant $U$. The main point to note is that the even-odd effect  (sign-change for the current) appears only provided that $U$ 
exceeds a threshold value. Below this threshold, the critical current displays no change of sign with increasing $L$, although there are small oscillations in the quantity 
superimposed on a monotonic decay. However, increasing $U$ sufficiently will induce $0-\pi$-transitions for odd junctions, while only decreasing the effective interface 
transparency for even junctions, as seen in Fig.~\ref{fig:lattice}. Note that for $L=5$, the threshold value for observing a $0-\pi$-transition in the critical current is 
above $U=1.4$, while for $L=7$, the threshold value is below $U=1.4$. Since the magnitude of the staggered magnetic order parameter increases monotonically with $U$, this 
indicates that longer odd junctions require a weaker magnetization to undergo $0-\pi$ transitions. This result has some resemblance to the case of a ferromagnetic junction,  
where such transitions can occur regardless of the magnitude of the exchange field provided that the junction is sufficiently long. We will discuss this point in more detail 
below.

\begin{figure}
\centering
\includegraphics[width=\columnwidth]{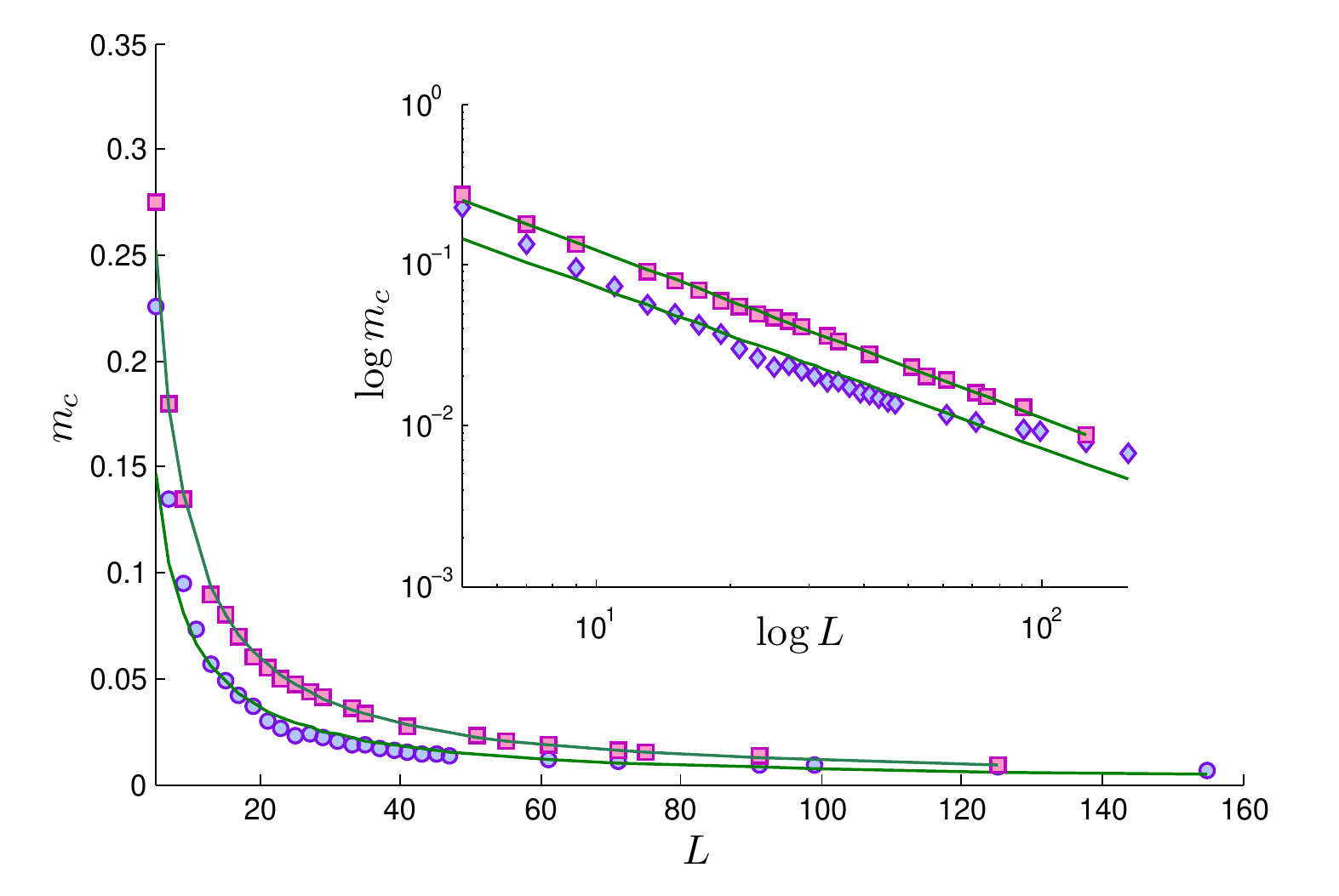}
\caption{(Color online) Critical value of the magnetic moment $m_c$ as a function of junction width $L$ for an antiferromagnetic junction (blue circles/diamonds) and a 
ferromagnetic junction (red squares). In the inset, the results plotted on a log-log scale for the ferromagnetic case fall on a straight line corresponding to $1/L$. The 
results for the antiferromagnetic case, however, saturate at large $L$. For the ferromagnetic case, the results show that for a wide enough SFS junction, any amount of 
ordering suffices to produce a $0-\pi$ transition in the junction. For the antiferromagnetic SAFS case, a threshold value is needed for this to occur, since only edge 
spins are uncompensated. We have used $V = 1.9$.}
  \label{fig:McL}
\end{figure}

In order to understand the physics behind these results, note that even junctions have a vanishing total magnetic moment as there is an equal number of spins with opposite 
direction. An even junction is essentially an SNS junction with the extra effect that increased staggering increases the resistance of the junction. This may be understood 
in simple terms by noting that an increasing staggering provides an increasing spin-dependent potential that scatters the current-carrying states in the AF-junction for low 
values of $U$.

Odd junctions, on the other hand, have a finite magnetic moment due to the uncompensated spin at one of the edges of the AF region. The layers adjacent to the interfaces have 
parallel spins for odd junctions, while even junctions have antiparallel spins. This resembles the situation in, for instance, SFIFS junctions with parallel or anti-parallel 
alignment of the exchange fields of the two ferromagnets,~\cite{barash_prb_02} or SIFIS junctions where the interfaces are spin-active.~\cite{enoksen_prb_12} In odd junctions, 
the subgap states are spin-split, while for even junctions, the states are spin degenerate. Naively, odd junctions are thus equivalent to ferromagnetic junctions, as there 
is a net magnetic moment (albeit weak) if one averages over the junction length. One notable difference is, however, that the average magnetic moment density in the SAFS-case 
scales with $1/L$, while it is constant in the SFS-case. Moreover, it is known in the SFS case that, provided the junctions are wide enough, any finite amount of magnetic 
moment density suffices to induce a $0-\pi$ transition. For the SAFS case, it is far from obvious that one should get the same behavior, due to the effective $1/L$-scaling of 
the average magnetic moment density. It is therefore of some importance to investigate the scaling with $L$ of the critical moment density required to induce $0-\pi$ oscillations 
in the SAFS-case, and compare it with the behavior of the SFS-case.  
 
To do so, it is instructive to consider an analogy with a ferromagnetic junction, where the Andreev-bound state energy in the ballistic limit reads~\cite{cayssol_prb_04}: $\varepsilon_{\sigma}(\varphi) = \Delta \left|\cos{\left(\frac{\varphi + \sigma\alpha}{2} \right)} \right|.$ Here, $\alpha = \frac{2 E_{ex}L}{\hbar v_F},$ 
where $E_{ex}$ is the ferromagnetic exchange energy, and $v_F$ is the Fermi velocity. We have assumed that the interface transparency is perfect in order to use a simple 
analytical expression for the ensuing discussion. Now, for even $L$ the total magnetic moment of the antiferromagnet is zero, hence we get no $0-\pi$ transition in even 
junctions. For the odd junction in question we can relate the finite magnetic moment to an effective ferromagnetic exchange field. The two factors contributing to this 
extra phase is the magnetic moment and the width of the junction, by analogy to an effective ferromagnet for odd $L$. A long junction can display a $0 - \pi$ transition 
for a smaller magnetic moment magnitude than a short junction, as seen in the right panel of Fig.~\ref{fig:lattice}. For $U = 1{.}4$, $L = 5$ has not undergone the 
transition yet, while it has undergone a transition for  $L=7$ and longer junctions. This, superficially at least, seems to indicate that the behavior is similar to that 
of an SFS-junction. Note however, that in the SFS-case, the parameter $\alpha$ increases linearly with $L$ and will exceed any prescribed threshold value, provided $L$ 
is large enough. For an SAFS-junction, however, where one naively expects $E_{ex} \sim 1/L$ such that $E_{ex} L \sim 1$, a threshold value for 
the magnitude of the staggered order parameter would be required in order to observe a $0-\pi$ oscillation.  

To investigate this further, we have numerically compared an SFS-junction with an odd SAFS-junction. Fig.~\ref{fig:McL} shows the critical magnetic moment as a function of 
junction width for both kinds of junctions, SFS and SAFS. Here, the critical magnetic moment is defined as the value where the critical current changes sign for a given 
value of the junction length $L$. Although the fits appear to be quite similar on a linear scale, they are significantly different when viewed on a log-log scale. The 
results for the SFS-junction follows the intuition one gets based on the bound-state energy and phase-shifts shown above, namely that the critical value of the magnetic 
moment density, which is proportional to a threshold value of $E_{ex}$ in order to get a $0-\pi$ transition, scales as $1/L$. This is not so for the SAFS-case. Rather, 
since $E_{ex} \sim 1/L$ in this case, one expects the critical value of the magnetic moment density to be non-zero also for long junctions. This is indeed 
seen in the log-log plot, where the results for the SFS-junction falls on a linear curve, while the results for the SAFS-junction reaches a constant value asymptotically.   

\begin{figure}[t!]
  \centering
  \includegraphics[width=\columnwidth]{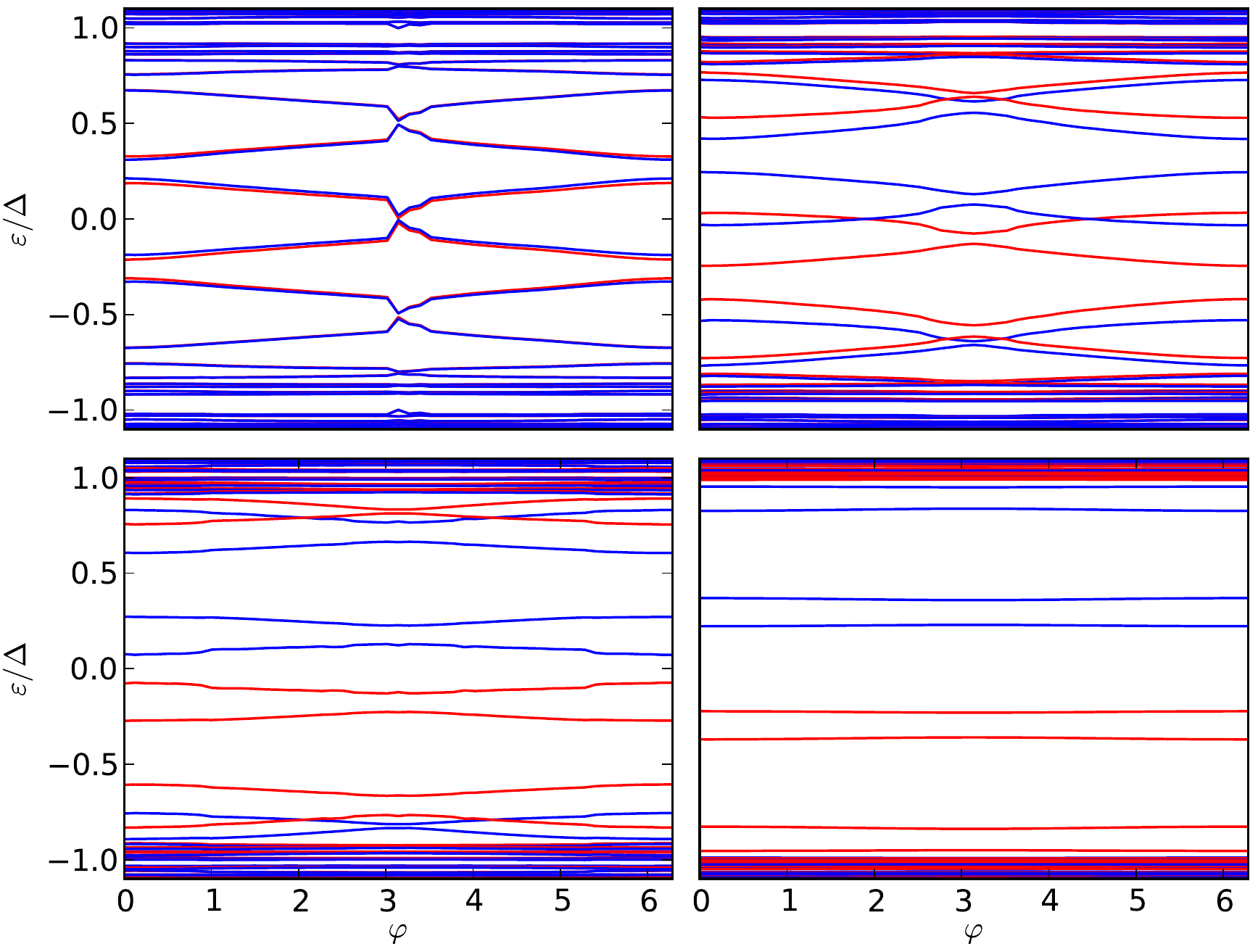}
  \caption{(Color online) Energy-levels of current-carrying states of the system as a function of superconducting phase-twist across the junction. Red lines denote spin-up 
states, blue lines denote spin-down states. As the strength of the antiferromagnetic ordering increases ($U$ increases), spin-down states are lowered in energy by the 
Zeeman-effect due to the uncompensated spin, while spin-up states increase in energy. Upper left panel: $U=0.5$. Upper right panel $U=1.2$. Lower left panel: $U=1.545$. Lower 
right panel: $U=2.0$ The contributions to the current are essentially determined by the $\phi$-derivative of each of the curves. As the levels cross, the levels contributing 
most significantly to the current changes sign, leading to the $0-\pi$ transition. This requires a threshold value, unlike in the ferromagnetic case.  Other parameter values 
are: $V=1.9$, $\mu=0$, $L=5$.}
\label{fig:Tableau}
\end{figure}

Furthermore, we have considered the crossing of the energy-levels of the current-carrying states 
as the strength of the antiferromagnetic ordering, parametrized by $U$, is increased. This is shown in Fig. \ref{fig:Tableau}. Level crossings of spin-up states and spin-down 
states, with opposite $\phi$-dispersion, occur for large enough value of $U$, i.e. as the magnitude of the magnetic ordering increases. Such level crossings tend to reverse 
the sign of the current, which 
is essentially determined by the $\phi$-derivative of the levels. The levels that contribute most significantly to the currents are seen to be the levels close to zero energy, as 
the levels further away are $\phi$-independent and therefore carry little current. The main difference between such spectra for the SAFS-junctions and the corresponding ones for
SFS-junctions, is that complete level-reversion of near-zero energy states occurs much more easily in the SFS-case than the SAFS case, since the total magnetic moment scales 
with the width of the junction in the SFS-case, while it does not in the SAFS-case. Hence, the energy bands with a definite spin-content are much more susceptible to a Zeeman effect 
in the SFS-case, compared to the SAFS-case.  In particular, one feature of the SAFS-spectra shown below is that there is essentially only one level-reversion between spin-up and spin-down subgap states before the bands flatten out, thus no longer contributing to the currents. For the SFS-case (not shown here), there are several level-inversions of subgap-states with only minor flattening of the bands as the magnetization increases, thus providing a much more efficient way of reversing the sign of the currents over the junction.  

{It is clear that antiferromagnetic order is not in itself sufficient to induce $0-\pi$ transitions in the supercurrent.} Ref. \onlinecite{andersen_prl_06} showed that for a 
sufficiently strong staggered order parameter, the current-phase relation of an antiferromagnetic Josephson junction revealed a $0$- or $\pi$-state depending on whether the 
antiferromagnet had an even or odd number of atomic layers. {We have above explicitly studied the actual threshold value of the magnitude of the staggered order parameter 
where transitions cease to occur}, regardless of whether the interlayer has an odd or even number of lattice sites. To observe the even-odd effect (and its absence as predicted 
here for weak antiferromagnets), it would be necessary to exert satisfactory control of the junction length (of order lattice spacing). It would certainly be an experimental 
challenge to tailor the junctions in this way~\cite{robinson_prl_10}. Nevertheless, previous studies have demonstrated that it is possible to fabricate ultrathin films with 
atomic-scale control over the thickness \cite{guo_science_04, eom_prl_06}, which would allow for a test of our predictions. It is difficult within the framework of our model to make quantitative predictions for actual parameter values in candidate systems. That would require a more refined approach including non-ideal effects such as disorder both in terms of non-magnetic impurities and with regard to the magnetic sublattices. 

The threshold value predicted here has an interesting consequence, namely that it becomes possible to \textit{induce a $0-\pi$ transition in antiferromagnets via pressure}. The 
reasoning is as follows. For an odd-$L$ junction with sufficiently strong AF order, the junction is in the $\pi$-state. Were one to bring the magnitude of the AF order 
back down below the threshold value for the $0-\pi$ transition, one would effectively have reversed the dc Josephson-current across the junction. Suppression of the AF 
order may effectively be obtained by destroying the Fermi-surface nesting, something that can be achieved by applying pressure to the sample. The pressure may be induced 
via an electric mechanism, such that one effectively is using electric means to control magnetic order, and by that in turn current-switching. This effect has no counterpart 
in an SFS-junction, since ferromagnetism does not rely on Fermi surface nesting. The combination of a threshold value for the staggered order 
parameter at which one may have $0-\pi$ oscillations, and the possibility to destroy nesting and thus AFM order via pressure, amounts to electrically controlled $0-\pi$ 
transitions in a single sample. This sets the results for an SAFS-junction apart from what is found for the ferromagnetic SFS-case, where any amount of magnetic order 
in principle suffices to induce $0-\pi$ transitions. Such transitions typically occur by changing the junction length, requiring multiple samples (although 
temperature-dependent transitions are also possible). The AFM order thus offers a novel mechanism for controlling the quantum mechanical ground state of a Josephson 
junction.

H.E. thanks NTNU for financial support. JL and AS acknowledge support from the Norwegian Research Council, through Grants 205591/V20 and 216700/F20.

\end{document}